\documentclass{aa}

\usepackage{graphicx}
\usepackage[varg]{txfonts}
\usepackage[colorlinks=true, allcolors=blue]{hyperref}

\usepackage{amsmath}
\usepackage{amssymb}
\usepackage{textcomp}
\usepackage{gensymb}
\usepackage{mathtools}
\usepackage{subcaption}
\usepackage{multirow}
\usepackage{colortbl}
\usepackage[dvipsnames]{xcolor}
\usepackage{arydshln}

\makeatletter
\renewcommand*\aa@pageof{, page \thepage{} of \pageref*{LastPage}}
\makeatother

\begin{document}

\title{Pulsar J1909+0749 and Supernova Remnant G42.1-0.2}
\subtitle{A case of a pulsar that has exited the shell of its associated supernova remnant}
\titlerunning{Speedy}

\author{
    R.\,Dokara \inst{1}
    \and
    M.\,Kramer \inst{1}
    \and
    A.\,Brunthaler \inst{1}
    \and
    M.\,J.\,Keith \inst{2}
    \and
    B.\,W.\,Stappers \inst{2}
}
\authorrunning{Dokara et al.}

\institute{Max Planck Institute for Radioastronomy (MPIfR), Auf dem H\"ugel 69, 53121 Bonn, Germany\\
           \email{rdokara@pm.me}
           \and
           Jodrell Bank Centre for Astrophysics, University of Manchester, Manchester M13 9PL, UK
           }

\date{Received 28 July 2026; accepted 19 October 2026}



\abstract
{ The ATNF pulsar catalogue currently features over 4300 Milky Way pulsars, but only 63 of them are possibly associated with a supernova remnant (SNR) arising from the same progenitor supernova explosion.  Even rarer is the scenario in which the pulsar has escaped the physical extent of its corresponding SNR. }
{ We aim to identify the serendipitously discovered shell-type object G42.1-0.2 as a SNR.  The known pulsar (PSR) J1909+0749 lies well outside the shell of SNR G42.1-0.2, but it has a cometary tail pointing towards the centre of the SNR; because of this positional coincidence, we propose that the SNR and the pulsar originated from the same supernova explosion. }
{ By measuring the flux densities of the shell G42.1-0.2 from ${\sim}$200~MHz to ${\sim}$6~GHz, we establish the nonthermal nature of its emission.  With the characteristic age of the pulsar, its distance from the Sun estimated from the dispersion measure (DM), and the angular separation between the pulsar and the centre of the shell, we infer the evolutionary state of the SNR and the transverse velocity of the pulsar.  Using proper motion measurements, we compare the SNR-pulsar direction with the velocity of the pulsar.  }
{ The radio spectral index of the shell is $-0.36 \pm 0.08$, confirming that the emission is primarily nonthermal synchrotron.  The proper motion measurements are not well constrained, but the amplitude and the direction are consistent with the interpretation that the SNR and the pulsar are associated.  If the characteristic age and the DM-based distance represent the true age and the true distance of the pulsar and the SNR, then the standard evolution equations suggest that the SNR must still be in the Sedov-Taylor phase, evolving in a medium with a density of about 1.5~cm$^{-3}$.  In addition, as a result of the PSR-SNR association, the transverse velocity of the pulsar PSR\,J1909+0749 is estimated to be \textit{${\sim}1460~\mathrm{km\,s}^{-1}$}, ranking it among the highest transverse velocities observed for pulsars in the Milky Way.  }
{}


\maketitle

\begin{figure*}[ht]
    \centering
    \includegraphics[width=0.93\textwidth]{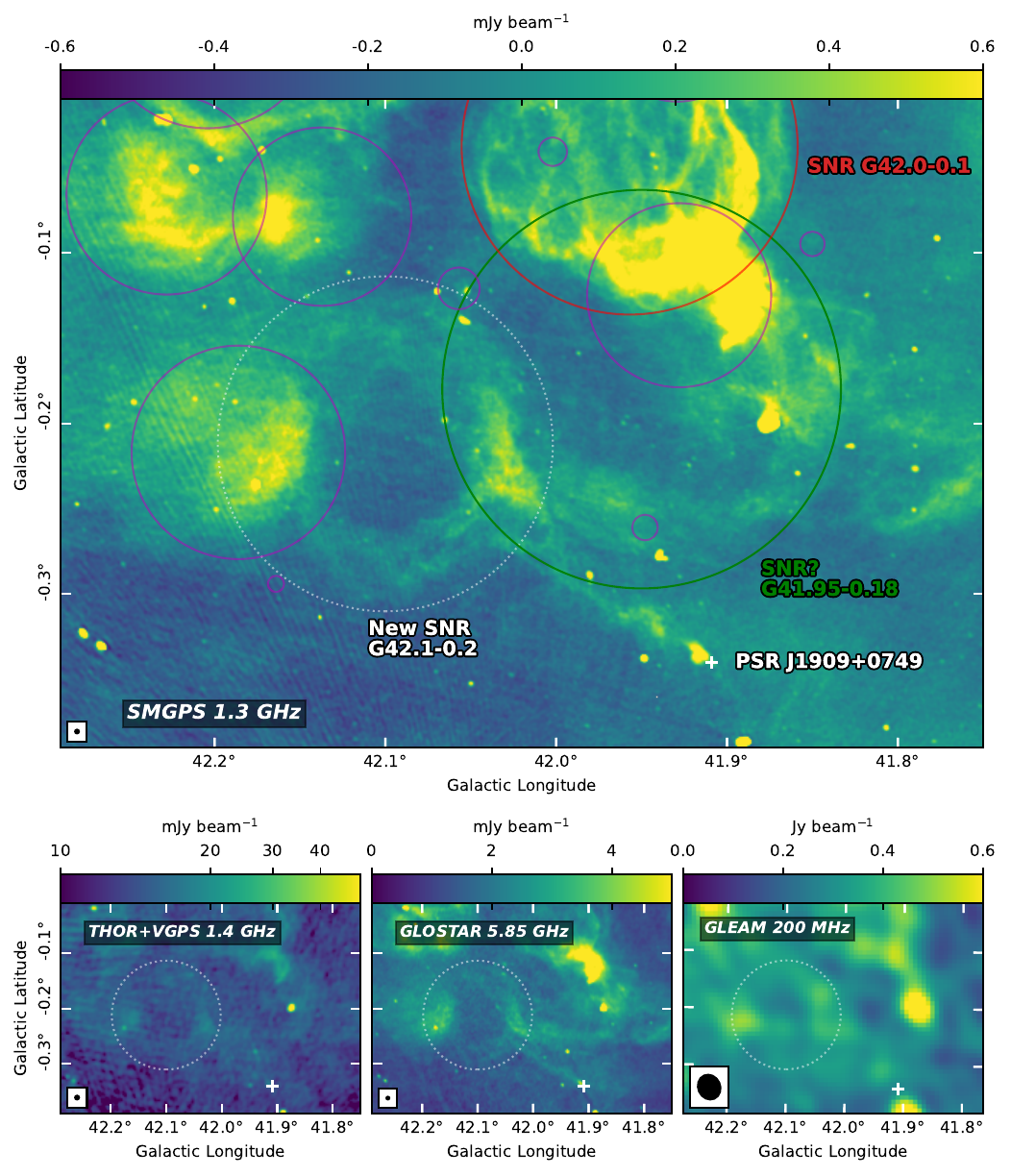}
    \caption{Continuum emission from the pulsar PSR J1909+0749 \citep{2013ApJ...772...50N} and its surroundings, as seen in the SMGPS (top), the THOR+VGPS (bottom left), the GLOSTAR VLA-D+Effelsberg (bottom middle), and the 200~MHz GLEAM (bottom right).  In all the panels, PSR\,J1909+0749 and the new SNR G42.1-0.2 are marked with a white cross and a white dashed circle, respectively.  In the top panel, a few additional objects are marked: the known SNR G42.0-0.1 \citep{2002ApJ...566..378K} in black, the SNR candidate G41.95-0.18 \citep{2017A&A...605A..58A} in green, and the H\,{\footnotesize II}~regions from the WISE catalog \citep{2014ApJS..212....1A} in magenta.}
    \label{fig:main}
\end{figure*}

\begin{table*}[ht]
\caption{Continuum data used in this study.}
\label{tab:datasources}
\centering
\begin{tabular}{l c c c c c}
\hline\hline
Data  &  Instrument(s)  &  Frequency  &  Resolution  &  LAS  &  References \\
      &              &    (GHz)    &    ($''$)    & ($'$) &    \\
\hline\\[-2mm]
GLEAM & Murchison Widefield Array &  0.07--0.23 &  ${\sim}$120 & ${\sim}$1740 & [1] \\
SMGPS & MeerKAT &  0.9--1.7   &  8           & ${\sim}$30   & [2] \\
THOR+VGPS & VLA[C+D] + Effelsberg &  1--2  &  25          & ---          & [3] \\
GLOSTAR & VLA[D] + Effelsberg & 4--8 & 18  & ---     & [4] \\
\hline
\end{tabular}
\tablefoot{
    [1] \citet{2015PASA...32...25W} and \citet{2019PASA...36...47H}.  
    [2] \citet{smgps}.
    [3] \citet{thor2016}, \citet{thor2020}, and \citet{vgps2006}.
    [4] \citet{medina2019}, \citet{brunthaler2021}, and \citet{glostar7}.
}
\end{table*}

\section{Introduction}

The ejected matter from a supernova (SN) explosion expands and sweeps up the material in the surrounding interstellar medium (ISM), forming what we call a supernova remnant (SNR).  SNRs inject momentum and energy into the ISM and enhance the turbulent gas pressure and the metallicity \citep[e.g.,][]{2005ARA&A..43..337C,dubner2015,2016ApJ...826..189M,2017ApJ...834L...3Y,2020ApJ...905...35K}.  While the feedback effects due to SNRs and their progenitor SN explosions on star-formation are routinely included---either directly or indirectly---in high-resolution galaxy-scale simulations \citep[e.g.,][]{2012MNRAS.423.1726S,2023MNRAS.523.3709C}, the initial conditions are poorly constrained as of yet, making the simulations overly sensitive to the initial conditions \citep[e.g.,][]{2018ApJ...861..107L,2024ApJ...975..229Z}.  This is partly due to the challenges in determining the physical parameters of SNRs through observations.  

Notably, determining the distance to a SNR reliably is still a difficult task to achieve, despite its fundamental importance in estimating the SNR's size, expansion velocity, kinetic energy, age, and evolutionary state.  Although advances have been made recently \citep[e.g.,][]{2022ApJ...940...63R} in moving beyond the inadequate $\Sigma{-}D$ (surface brightness to distance) relationship \citep{2005MmSAI..76..534G}, the majority of SNRs still lack reliable distance measurements.

Exacerbating the situation is the fact that the current catalogs of SNRs in the Milky Way \citep{2012AdSpR..49.1313F,2025JApA...46...14G}, which contain fewer than 400 objects, are estimated to be at most ${\sim}30$\% complete \citep{1991ApJ...378...93L,2022ApJ...940...63R}.  In essence, our knowledge of the physical parameters of Galactic SNRs comes from an under-sampled population whose distances are reliable only for a minority, potentially potentially distorting our understanding of their true Galaxy-wide characteristics.

A likely byproduct of a SN event of the core-collapse (CC) type, which is responsible for at least 70\% of the SN events \citep[e.g.,][]{1994ApJS...92..487T,2006Natur.439...45D}, is a rapidly rotating neutron star known as a pulsar \citep[PSR; e.g.,][]{2003ApJ...591..288H}.  A characteristic age of the pulsar, which may be used as a proxy for its true age \citep[see, e.g.,][for caveats]{2000Natur.406..158G}, can be derived using the time-period and its derivative.  In addition, the distance to the pulsar can be inferred using the dispersion measure (DM) and an electron density model of the Galaxy \citep[e.g.,][]{2017ApJ...835...29Y}.  Consequently, if a pulsar is associated with a SNR, then the age and the distance of the remnant can be regarded as those of the pulsar.

However, SNR-PSR associations are not commonly found.  According to the ATNF pulsar catalogue \citep{2005AJ....129.1993M}, there are only 63 possible associations of pulsars and SNRs in the Milky Way at the moment.  Of these, only 44 associations are considered to be confirmed.  The low number of confirmed SNR-PSR associations is because of two reasons.  First, pulsars may continue radiating for millions of years, whereas SNRs have a much shorter mean lifetime of the order of $10^5$~yr \citep{1994ApJ...437..781F}.  Second, pulsars can have kick velocities as high as ${\sim}1000~\mathrm{km\,s}^{-1}$ \citep{1993Natur.362..133C,2009ApJ...698..250C}, which may be large enough for the pulsar to escape even its corresponding SNR shell.  The Mouse \citep{1987Natur.330..455Y} is one well-known example of such a scenario.  Recently, a similar SNR-PSR association has been discovered by MeerKAT, which was named the Mini Mouse \citep{2023MNRAS.523.2850M}.  Only a handful of such cases are known in which the pulsar escaped the shell of its sibling SNR: e.g., the Mushroom \citep{2017JPlPh..83e6301K}, the Frying Pan \citep{1987AuJPh..40..855K,2009ApJ...703L..55C} and the Canon Ball \citep{2019ApJ...876L..17S}.

In this paper, we report one more such association in which the pulsar has escaped the confines of its SNR shell.  We discovered, by chance, a shell centred at the Galactic coordinates $l{\sim}42.1\degree$, $b{\sim}-0.2\degree$ with a diameter of ${\sim}11'$ and a cometary nebula pointing linearly away from the shell in the continuum images of the SARAO MeerKAT Galactic Plane Survey \citep[SMGPS;][]{smgps}.  The shell has not been identified by any study yet.  Furthermore, one corner of the cometary nebula is positionally coincident with the known pulsar PSR\,J1909+0749 (see Fig.~\ref{fig:main}).  We hypothesize that the shell is a SNR and that the pulsar was born in the same progenitor SN explosion as of the shell.  In Sect.~\ref{sec:data}, we describe the data and the methods used in this study.  In Sect.~\ref{sec:results}, we present the evidence supporting our hypothesis and discuss the implications of the SNR-PSR association in Sect.~\ref{sec:implications}.  We conclude with a short summary in Sect.~\ref{sec:summary}.

\section{Data and methods}
\label{sec:data}

\subsection{Radio continuum data}
\label{subsec:continuum}

We used radio continuum data from four surveys observing in the centimetre and metre wavelengths, which are summarised in Table~\ref{tab:datasources}.  The 4--8~GHz Global View on Star Formation (GLOSTAR) survey\footnote{\url{https://glostar.mpifr-bonn.mpg.de/glostar/}} includes data from the \textit{Karl G. Jansky} Very Large Array (VLA) in its B and D configurations and the 100-metre Effelsberg telescope \citep{brunthaler2021}.  In this study, we only use the short-spacing-corrected VLA-D + Effelsberg images which were made by feathering the D-configuration images with the Effelsberg images \citep{glostar7}.  We will hereafter refer to it as the `GLOSTAR data'.  The 1--2 GHz HI/OH/Recombination line survey\footnote{\url{https://www2.mpia-hd.mpg.de/thor/Data\_\%26\_Publications.html}} data \citep{thor2016} combined with the VLA Galactic Plane Survey data \citep[VGPS;][]{vgps2006}, which is called the `THOR+VGPS', is also flux-correct at large angular scales.  We also used the data from the GaLactic and Extragalactic All-sky Murchison Widefield Array survey\footnote{\url{http://gleam-vo.icrar.org/gleam\_postage/q/form}} \citep[GLEAM;][]{2019PASA...36...47H} and the SARAO MeerKAT Galactic Plane Survey\footnote{\url{https://doi.org/10.48479/3wfd-e270}} \citep[SMGPS;][]{smgps}.  The SMGPS and the GLEAM surveys do not correct for missing zero-spacing information.  This, however, does not affect our analysis of the shell since the largest angular scale recovered (LAS) in these surveys are significantly larger than the shell studied in this paper: LAS${\sim}30\degree$ in GLEAM and LAS${\sim}30'$ in SMGPS \citep{2019PASA...36...47H,smgps}.

\subsection{Flux density and spectral index}
\label{subsec:fluxmeas}

The radio spectral index, $\alpha$, is determined by a linear fit to the relationship between the log of the flux density, $S_\nu$, and the log of the frequency, $\nu$, i.e., $S_\nu \propto \nu^\alpha$.  The shells of SNRs typically have a spectral index of about $-0.5$, and obtaining a negative spectral index is imperative for an object to be considered a SNR \citep{2013tra..book.....W,2025JApA...46...14G}.  

However, for the spectral index to be calculated reliably, the contribution from the diffuse Galactic background emission to the flux density of the object must be completely removed.  Since the background is stronger at lower frequencies and also predominantly nonthermal at frequencies lower than 1~GHz \citep[e.g.,][]{2005MNRAS.360.1545P}, it can severely bias the results of the spectral index calculations towards negative values.  In order to subtract the background, we used the same method as in \citet{glostar7}.  A two-dimensional linear fit is performed on pixels containing only the background emission and the result of this fit is used to remove the locally varying background across the extended object.  

The uncertainty in the aforementioned subtraction technique is estimated by fitting multiple sets of background pixels and taking the spread among the resulting fits.  This component is then combined in quadrature with the uncertainty arising from the flux calibration scale and the local rms noise, yielding the total uncertainty in the flux density of an extended object.  The uncertainty in the spectral index, on the other hand, is derived through Monte-Carlo propagation of these flux density errors.  For each iteration, a synthetic dataset is generated in which every data point is drawn from a Gaussian distribution centered on its measured flux density, with a standard deviation equal to its corresponding uncertainty.  An ordinary least squares linear fit is performed on each synthetic dataset to obtain a trial value of the spectral index.  After many such realizations, the standard deviation of the resulting distribution of trial spectral indices is adopted as the uncertainty in the spectral index.

\section{Identification of G42.1-0.2 as a SNR}
\label{sec:results}

\subsection{Radio emission from SNRs}

SNRs emit nonthermal synchrotron radiation.  They are generally shell-shaped extended objects observed in radio wavelengths with negative spectral indices ($\alpha{\sim}-0.5$) and possibly linearly polarised \citep[e.g.,][]{2011A&A...536A..83S,2013tra..book.....W}.  In the SNR catalogs of \citet{2012AdSpR..49.1313F} and \citet{2025JApA...46...14G}, at least 60\% of the entries are noted to be shell-type, and ${\sim}90$\% are detected in radio.  

Alongside SNRs, H\,{\footnotesize II}~regions are the other major group of extended objects populating the Galactic plane.  They are also detected in radio, and are regularly confused to be SNRs \citep[e.g.,][]{2017A&A...605A..58A,2021A&A...651A..86D} due to the similar morphology of their radio emission.  H\,{\footnotesize II}~regions are easily detected in MIR due to the presence of dust and polycyclic aromatic hydrocarbons around young high-mass stars.  In contrast to SNRs, H\,{\footnotesize II}~regions emit thermally, have flat or rising spectra (spectral index $\alpha\gtrsim -0.1$), and are not inherently linearly polarised \citep[e.g.,][]{2013tra..book.....W}.  Another distinction between H\,{\footnotesize II}~regions and SNRs is that H\,{\footnotesize II}~regions emit strongly in both radio and MIR wavelengths, while SNRs are not readily observed in MIR.  This radio-MIR anti-correlation property has been widely used to separate these two classes of objects and also identify new SNR candidates \citep[e.g.,][]{1987A&AS...71...63F,1987Natur.327..211H,1996A&AS..118..329W,2006ApJ...639L..25B,2006AJ....131.2525H}.  Although the presence or absence of polarisation does not definitively distinguish between the nonthermal synchrotron radiation from SNRs and the thermal bremsstrahlung radiation from H\,{\footnotesize II}~regions \citep[e.g.,][]{2018ApJ...866...61D,2021A&A...651A..86D}, the radio spectral index and the MIR emission serve as strong indicators.

\subsection{The shell: G42.1-0.2}

We identified the radio shell G42.1-0.2 and a linear, cometary feature pointing away from it (with the pulsar PSR\,J1909+0749 at the end of this feature) first in the 1.3~GHz SMGPS data (Fig.~\ref{fig:main}).  Presumably due to its low surface brightness and the complex surroundings, it has not been named as a candidate in the growing literature on new SNR candidates discovered in this region \citep[e.g.,][]{2025A&A...693A.247A,2024A&A...690A.247T,2021A&A...651A..86D,2019PASA...36...45H,2017A&A...605A..58A}.  

The region surrounding this newly identified shell contains the known SNR G42.0-0.1 \citep[][]{2002ApJ...566..378K,2012AdSpR..49.1313F,2025JApA...46...14G} and the SNR candidate G41.95-0.18 \citep{2017A&A...605A..58A}, in addition to numerous H\,{\footnotesize II}~regions and unresolved sources.  The shell G42.1-0.2 overlaps with the SNR candidate G41.95-0.18 on the west and the H\,{\footnotesize II}~region G042.186-00.218 \citep{2014ApJS..212....1A} on the east.  This H\,{\footnotesize II}~region is fully excluded from the flux density measurements of G42.1-0.2.  The SNR candidate G41.95-0.18 was visually identified in the THOR+VGPS data by \citet{2017A&A...605A..58A} using the radio-MIR anti-correlation property.  This SNR candidate has significant overlap with G42.1-0.2.  However, due to the low detection significance of the rest of the shell of the candidate G41.95-0.18 and the relatively more coherent appearance of the shell of G42.1-0.2 in the SMGPS and the GLOSTAR images, we consider the overlapping region to be entirely a part of the shell of G42.1-0.2.

\begin{figure}[t]
    \centering
    \includegraphics[width=0.45\textwidth]{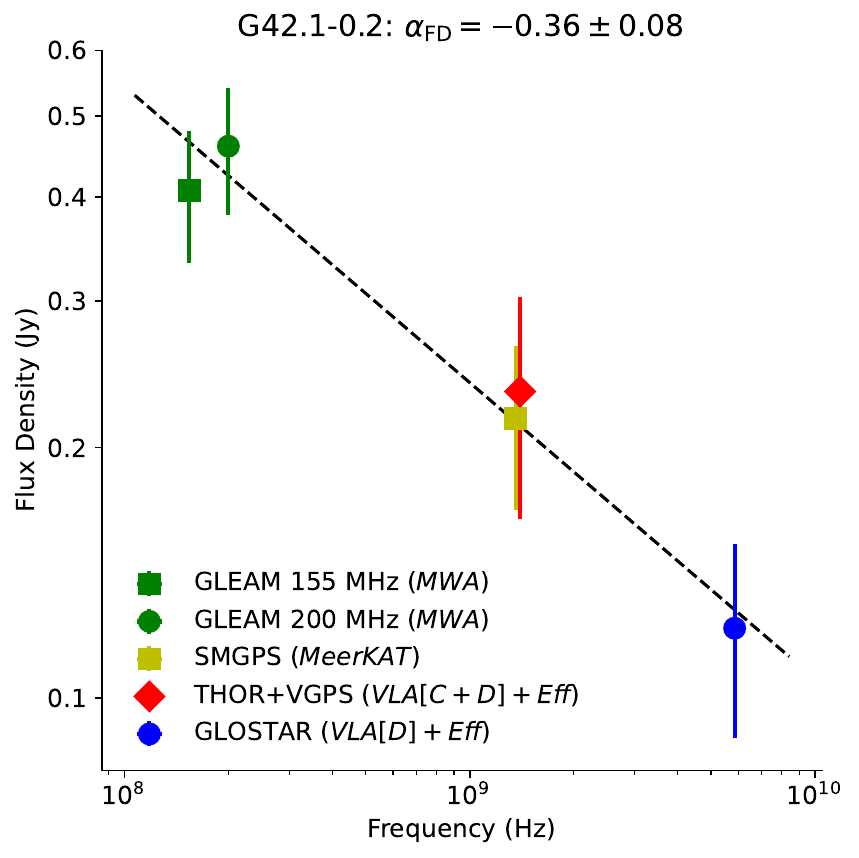}
    \caption{Radio spectrum of G42.1-0.2.  The methods to measure the flux densities and to estimate the uncertainties in the flux density and the spectral index are described in Sect.~\ref{subsec:fluxmeas}.}
    \label{fig:shellspectrum}
\end{figure}

The flux densities of G42.1-0.2 are measured using the images of the 5.85~GHz GLOSTAR, 1.3~GHz SMGPS, 1.4~GHz THOR+VGPS, and the 155/200~MHz data of the GLEAM survey.  We only used the two high-frequency bands of the GLEAM survey data because of their relatively better resolution and lesser source confusion\footnote{In addition, free-free and synchrotron self-absorption effects significantly decrease the flux density below ${\sim}100$~MHz \citep[e.g.,][]{2013tra..book.....W}.}.  As mentioned in Sect.~\ref{subsec:fluxmeas}, the local background is subtracted in all the images.  The values of the flux densities are presented in Table~\ref{tab:FDs} and the flux density spectrum of G42.1-0.2 is shown in Fig.~\ref{fig:shellspectrum}.  

\begin{table}[t]
\caption{Flux densities of G42.1-0.2.}
\label{tab:FDs}
\centering
\begin{tabular}{c c}
\hline\hline
Frequency   &  Flux density \\
  (GHz)     &    (Jy)    \\
\hline\\[-2mm]
0.155       &  $0.41\pm 0.07$ \\
0.200       &  $0.47\pm 0.09$ \\
1.3         &  $0.22\pm 0.05$ \\
1.4         &  $0.23\pm 0.07$ \\
5.85        &  $0.12\pm 0.03$ \\
\hline
\end{tabular}
\end{table}

We calculated the spectral index to be $-0.36 \pm 0.08$, which is evidence for nonthermal synchrotron emission.  Furthermore, in the $8~\micro$m GLIMPSE (Galactic Legacy Infrared Mid-Plane Survey Extraordinaire) and the $24~\micro$m MIPSGAL (Multiband Infrared Photometer for Spitzer Galactic plane survey) images \citep{2009PASP..121..213C,2009PASP..121...76C}, which have sensitivities capable of detecting continuum emission from all H\,{\footnotesize II}~regions in the Milky Way \citep{2014ApJS..212....1A}, we do not find any significant emission that corresponds to the shell, reasserting that G42.1-0.2 is a SNR.  While we did not detect any linear polarisation from the shell in the GLOSTAR data, this is quite common in SNRs and not unexpected \citep[e.g.,][]{2011A&A...536A..83S,2018ApJ...866...61D,2021A&A...651A..86D}.

\section{PSR J1909+0749}
\label{sec:implications}

\subsection{Possible Pulsar Wind Nebula (PWN)}

The pulsar PSR J1909+0749 was first reported in the work of \citet{2013ApJ...772...50N}.  We used updated timing information obtained with the Lovell telescope \citep[see e.g.,][]{2022MNRAS.512.2446N} to estimate the proper motion, i.e.~PMRA$=3\pm35$ mas yr$^{-1}$ and PMDEC$=-50\pm60$ mas yr$^{-1}$. Although these values are consistent with zero, we accept them at face value in order to investigate the possible direction and amplitude required for an association. 

The NE2025 electron density model  \citep{2026ApJ..1002....3O}  predicts a distance of 9.7 kpc based on the measured dispersion measure. 
The pulsar's spin-down yields a characteristic age of only $\tau_c = 24700$ years. This young age is consistent with the observation of a glitch in this pulsar \citep{2024MNRAS.532..859L}.  The surface magnetic field quoted in the ATNF pulsar catalogue \citep{2005AJ....129.1993M} is nearly $6\times 10^{12}$~G, which is relatively strong and indicative of the youth of the pulsar (see also Appendix~\ref{apdx:bsurf}).

Interestingly, we detect significant nebulous radio continuum emission immediately following pulsar PSR J1909+0749 (see Fig.\,\ref{fig:main} top panel).  Using the GLOSTAR and the SMGPS images, we measured the flux density of this nebula to be $6.6 \pm 3.0$~mJy and $15.1 \pm 2.7$~mJy, respectively.  The region used to measure the flux densities is marked in Fig.~\ref{fig:centrepoints}.  This gives a spectral index of $-0.6 \pm 0.4$, indicating non-thermal emission, although the large uncertainties mean that a flatter spectrum cannot be ruled out.

A Pulsar Wind Nebula (PWN) is the nebulous emission resulting from the relativistic winds powered by the pulsar.  They typically have a `flat' spectrum in radio wavelengths with $-0.3\,{\lesssim}\,\alpha\,{\lesssim}\,0$ \citep[e.g.,][]{1978A&A....70..419W,2006ARA&A..44...17G}, but spectral breaks and relatively steep spectra are also not uncommon (e.g., DA\,495: \citealt{2008ApJ...687..516K}; G29.38+0.10: \citealt{glostar7}).  It appears that this cometary emission is the PWN associated with the pulsar.

Furthermore, in the SMGPS images, a linear trail can be identified by connecting the pulsar's position, the cometary nebula's direction, and the centre of SNR G42.1-0.2.  Such a feature, routinely called the `tail' of the PWN, forms due to the interaction of the PWN's relativistic wind with the surrounding ISM.  As the pulsar moves supersonically, a bow-shock develops ahead, while the wind that lags behind creates an elongated tail that is significantly larger than the compact nebula \citep[e.g.,][]{1987Natur.330..455Y,2023MNRAS.523.2850M}.  The emission from the tail is expected to be nonthermal synchrotron due to the relativistic electrons gyrating in the magnetized plasma downstream of the termination shock \citep[e.g.,][]{2008AIPC..983..171K}.  Due to energy-dependent synchrotron cooling, higher-energy X-ray-emitting electrons have shorter lifetimes (order of decades to centuries, depending on the magnetic field) than the lower-energy electrons producing radio emission, which have typical lifetimes of 10~kyr to 100~kyr in the Milky Way's ISM.  PWN tails can thus be detected over larger scales of time and distance in radio than in X-rays \citep[see Sect.~5 of][for a review of the observations of PWN tails]{2017JPlPh..83e6301K}.  These tails are often faint compared to the compact nebula due to weakening magnetic fields in the downstream flow\footnote{Note that magnetic field weakening can be avoided or even reversed if the wind flow decelerates rapidly, as observed in PSR\,J1509-5850 \citep{2016ApJ...828...70K}, where radio surface brightness first increases with distance from the pulsar before falling off.}.  The tail of PSR\,J1909+0749 is detected in both the SMGPS (flux density $32 \pm 7$~mJy) and the GLOSTAR (flux density $11 \pm 5$~mJy), allowing us to estimate its spectral index of ~$-0.8 \pm 0.5$, indicating nonthermal emission.  The region used to measure the flux densities of the tail can be seen in Fig.~\ref{fig:centrepoints}.

Although the direction of the pulsar tail does not always trace the pulsar's proper motion \citep[e.g.,][]{2017ApJ...844....1K}, the visual alignment and the radio spectra lead us to propose that the pulsar and the SNR are the result of a single supernova explosion.

\subsection{Implications of the SNR-PSR association}

In order to derive the physical parameters of the SNR, we make the following initial assumptions:
\begin{itemize}
    \item The characteristic age of the pulsar is its true age.
    \item The DM-based distance to the pulsar as derived by the NE2025 model is correct and both the SNR G42.1-0.2 and the pulsar PSR\,J1909+0749 are at this distance.
    \item The pulsar and the SNR arise from the same progenitor SN explosion.
    \item The progenitor SN explosion took place at the centre of the SNR G42.1-0.2 ($l{\sim}42.097\degree$, $b{\sim}-0.229\degree$, see Appendix~\ref{apdx:snrcentre} for details of the measurement).
\end{itemize}

We estimate that the physical diameter of the SNR G42.1-0.2 is ${\sim}$33.2~pc using the pulsar's DM-based distance.  According to the standard SNR evolution equations \citep[e.g.,][]{2011piim.book.....D}, for a SNR to have a diameter of 33.2~pc at the age of 24.7~kyr, it must have evolved in an ambient medium that has a number density of ${\sim}$1.5~cm$^{-3}$, assuming that the initial explosion energy was $10^{51}$~erg.  In addition, the equations suggest that it is still in the Sedov-Taylor phase and that the snowplow phase would begin in another 10~kyr (when its age is 35~kyr).

\subsubsection{Proper Motion of the Pulsar}

\begin{figure*}[t]
    \centering
    \begin{tabular}{cc}
    \includegraphics[width=0.49\textwidth]{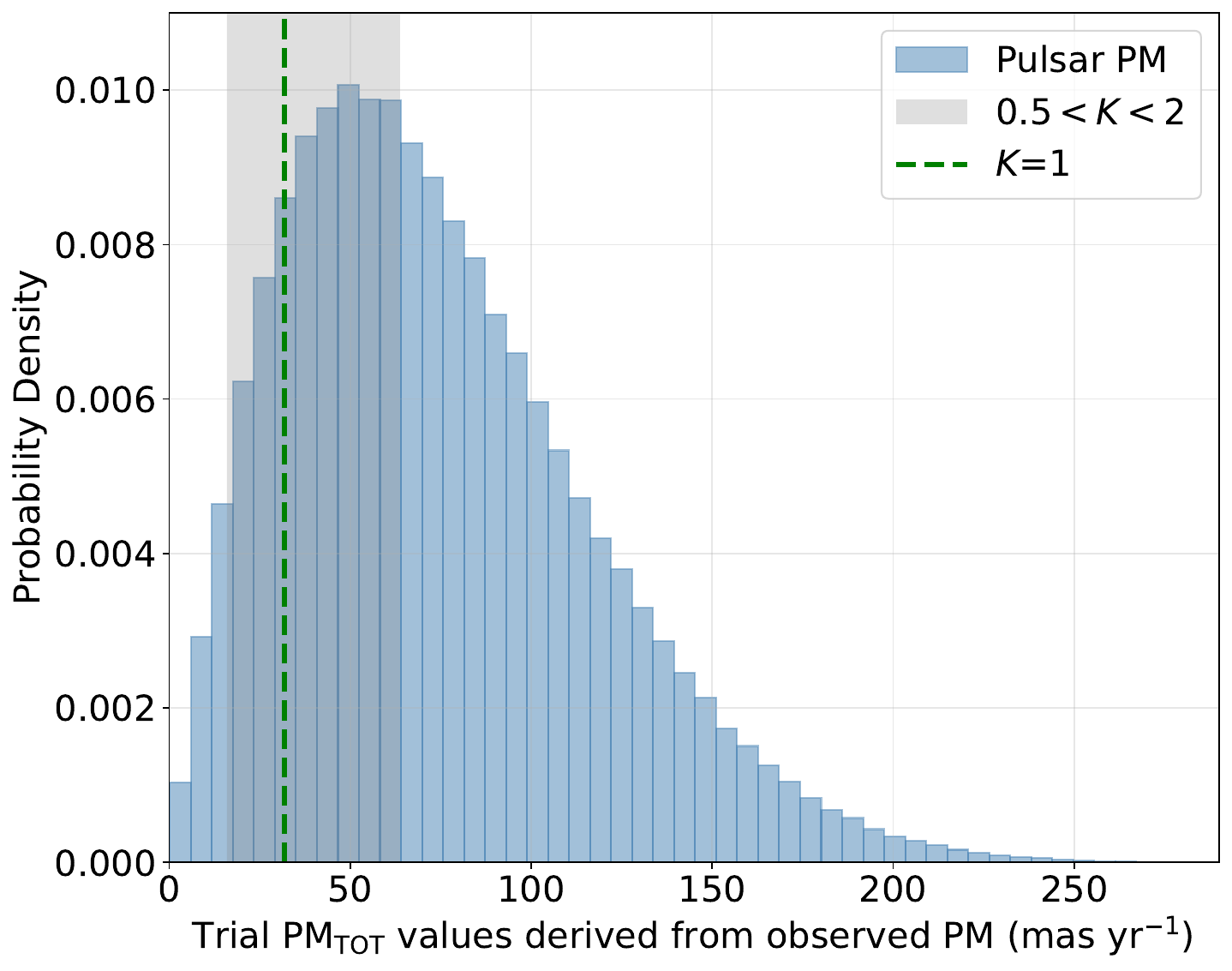} &
    \includegraphics[width=0.48\textwidth]{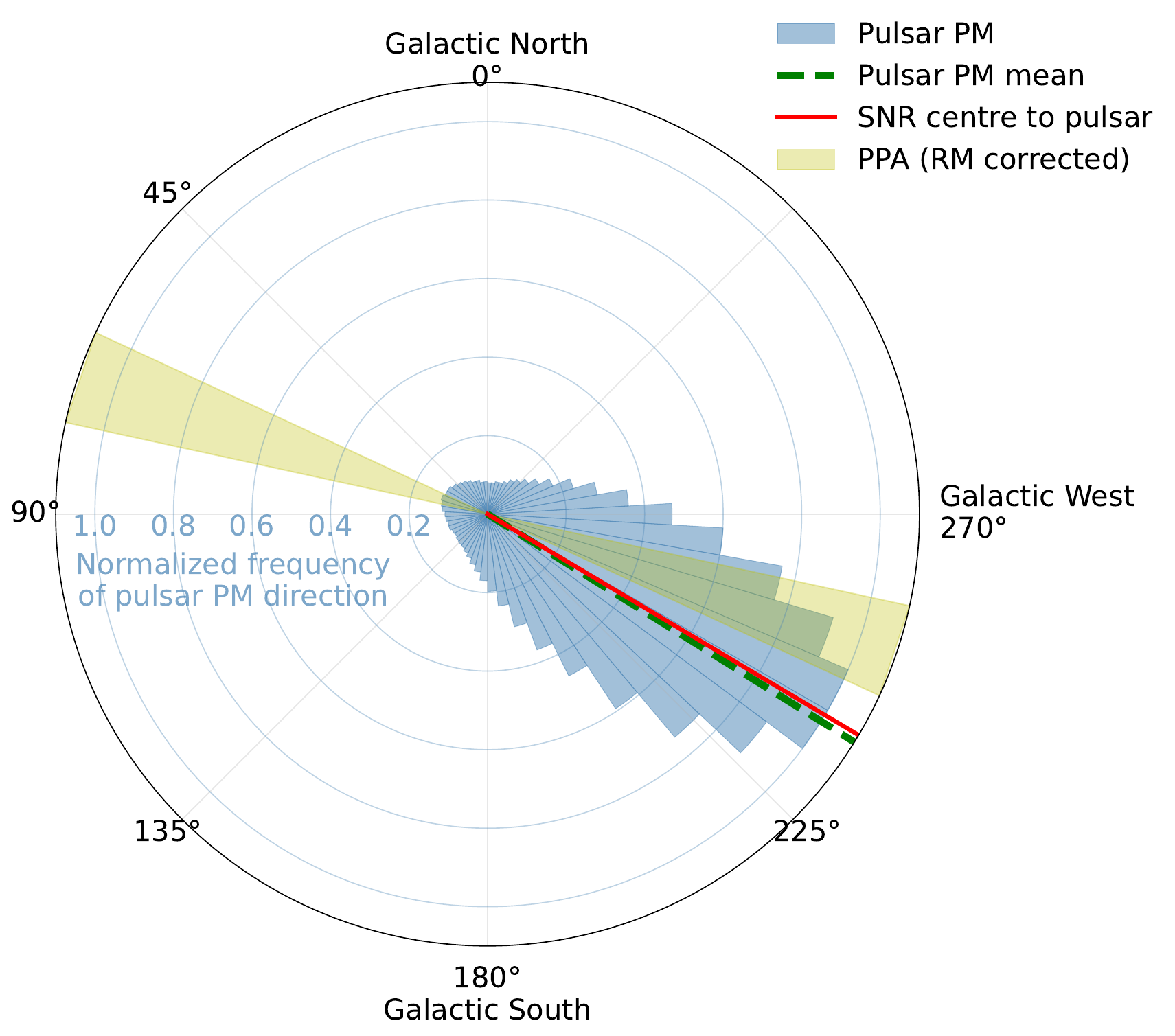} 
    \end{tabular}
    \caption{A comparison of the pulsar's proper motion (PM) with its separation from the SNR. (Left) Distribution of trial PM amplitudes obtained via Monte-Carlo sampling of the uncertainties derived from pulsar timing.  The shaded region indicates the amplitude required for the pulsar to have been born at the centre of the SNR, allowing up to a factor-of-two correction in its characteristic age (i.e., true age $\tau=K\times\tau_c$, with $0.5 < K < 2$).  The green vertical line marks the required amplitude assuming no correction, i.e., $K=1$.  (Right) Angular distribution of the directions of the trial PM vectors (blue wedges) is compared with the direction from the SNR centre to the pulsar, which is marked by a red line.  The circular mean of the angular distribution is marked with a green dashed line.  The direction of the pulsar spin, as implied by the absolute polarisation position angle (PPA), is indicated in yellow. See the text for details.
    }
    \label{fig:pm}
\end{figure*}

Given that the pulsar is located $13.1'$ away from the centre of the SNR, its proper motion estimated using the characteristic age is PMRA$=2\pm3$~mas\,yr$^{-1}$ and PMDEC$=-32\pm1$~mas\,yr$^{-1}$, which is consistent with the observed value of PMRA$=3\pm35$~mas\,yr$^{-1}$ and PMDEC$=-50\pm60$~mas\,yr$^{-1}$.  Using the DM-based distance, we estimate that the pulsar must have traveled a projected distance of 37~pc.  This gives the pulsar a transverse velocity of $\sim$1460~km~s$^{-1}$.
This is somewhat larger, but consistent with the largest transverse velocities measured for young pulsars (see \citealt{2019ApJ...875..100D} and Appendix~\ref{apdx:ps_tvels}).  

In Figure \ref{fig:pm}, we compare the amplitude and the direction of the observed proper motion with those required for the pulsar to have moved from the SNR centre to its current position within a time duration of the age of the pulsar, $\tau$.  A Monte-Carlo sampling is performed to generate synthetic datasets of PMRA and PMDEC from the observed values (assuming Gaussian errors), from which the trial values of the amplitude (PM$_{\mathrm{TOT}}$) and the direction in Galactic coordinates are obtained:

\begin{align*}
    &\mathrm{PM}_{\mathrm{TOT}} =
    \sqrt{\left(\mu_\alpha \cos\delta\right)^2 + \left(\mu_\delta\right)^2},\\
    &\theta_{\mathrm{PM}} =
    \arctan\!\left(\frac{\mu_\alpha \cos\delta}{\mu_\delta}\right) \bmod 360^\circ,
\end{align*}

where $\alpha$ is the right ascension, $\delta$ is the declination, $\mu_\alpha$ is the proper motion in right ascension, and $\mu_\delta$ is the proper motion in declination.  The histogram of the trial amplitudes is shown in the left panel of Fig.~\ref{fig:pm}.  The amplitude required for the pulsar to traverse this distance, given an age of $\tau=K\times\tau_c$, where $K$ is the correction factor with $0.5<K<2$, is indicated by the gray-shaded region.  
As can be seen, the observed proper motion is consistent with the estimated age of the pulsar, even if the characteristic age over- or underestimates the true age by a factor of two (or more).

The trial values of the direction of the PM vector are diagrammed as a histogram on a polar plot in Fig.~\ref{fig:pm} (right).  Since angles are wrapped around, in order to find the average direction, we used the \texttt{stats.circmean} method of the SciPy package \citep{2020SciPy-NMeth} to calculate the circular mean of the distribution.  This mean direction is indicated with a green dashed line.  The direction of movement from the SNR centre to the current pulsar position is shown with a red line in the figure, which is remarkably close to the mean direction of the current estimate of the proper motion, though we note that the uncertainties in the PM are quite large.  The overall direction is also consistent with the interpretation of seeing a cometary tail behind the pulsar.

\subsubsection{Proper Motion Direction of the Pulsar via Polarimetry}

Due to the significant uncertainty in the direction of the proper motion, we employ an alternative method to determine the pulsar's direction of motion. \cite{2005MNRAS.364.1397J} first demonstrated that the velocity vector of young pulsars is preferentially aligned with their spin vectors \citep[see also][and references therein]{2012MNRAS.423.2736N}. 
Hence, the position angle of the spin axis projected onto the sky may be treated as a proxy for the position angle of the velocity vector.  Conveniently, the spin axis can be estimated using the polarisation of the pulsar emission.  Pulsar radio emission is strongly linearly polarised, and the orientation of this polarisation, the polarisation position angle (PPA), sweeps smoothly through a characteristic S-shaped curve as the pulsar's beam sweeps across the observer's line of sight.  The centre of the swing corresponds to the plane defined by the spin axis and the observer's line of sight.  At this centre, the measured PPA equals the position angle of the spin axis projected onto the plane of the sky.  Hence, a measurement of the absolute PPA at the center of the swing acts as an indirect measurement of the direction of the pulsar's movement.  

We used MeerKAT observations of this pulsar \citep{2023MNRAS.520.4801J} to obtain the pulse profiles.  In order to recover the intrinsic PPA free of the wavelength-dependent Faraday rotation imposed by the interstellar medium, we corrected the Stokes $Q$ and $U$ parameters using the rotation measure RM$=-241\pm2$~rad\,m$^{-2}$ measured from these same data at 1281~MHz, yielding a PPA swing profile corrected to infinite frequency.  We then fitted the widely-used rotating vector model \citep{1969ApL.....3..225R} to the resulting swing.  Because the pulsar also emits a strongly polarised interpulse from its opposite magnetic pole, the fit is anchored at both poles simultaneously; this tightly constrains the pulse phase of the swing centre and removes the ambiguities that plague single-pole fits, permitting a robust estimate of the absolute PPA at the swing centre: PPA$=8.9^{\circ}\pm6.3^{\circ}$.  Both this direction and the complementary direction at $180^{\circ}+(8.9^{\circ}\pm6.3^{\circ})$, transformed into the Galactic coordinate reference system, are shown in Figure~\ref{fig:pm} (right).  Again, this is largely consistent with the SNR-pulsar direction.

\section{Summary}
\label{sec:summary}

We serendipitously discovered a shell-like object G42.1-0.2 and identified it as a new SNR due to its unambiguously nonthermal radio spectrum.  The pulsar PSR\,J1909+0749 was found to be outside the shell but nearby, with a cometary feature tailing towards the centre of the SNR G42.1-0.2.  We propose that these two stellar remnants originated from the same SN explosion and that the pulsar escaped the SNR, similar to the well-known `Mouse' PWN identified by \citet{1987Natur.330..455Y}.  The direction of the pulsar's velocity inferred from the measurements of the proper motion and the center of the PPA are consistent with this interpretation.  If the pulsar's derived age and distance are true, then the pulsar must have a transverse velocity of nearly 1460~km\,s$^{-1}$.  This inferred velocity would make it one of the fastest (in terms of transverse velocity) pulsars in the Milky Way.  Due to the PWN's apparent relative speed and similarities with the Mouse PWN, we nickname it `Speedy' after the cartoon character `Speedy Gonzales'.

\begin{acknowledgements}
    We thank Arshia M. Jacob for suggesting the nickname `Speedy' to the pulsar PSR\,J1909+0749.  The name `Speedy Gonzales' is used in this paper only as a nickname and is not affiliated with the studio Warner Bros.
    We made use of NASA's Astrophysics Data System, the SIMBAD database, and the softwares DS9 \citep{2003ASPC..295..489J}, Astropy \citep{2013A&A...558A..33A}, scikit-image \citep{2014PeerJ...2..453V}, and APLpy \citep{2012ascl.soft08017R} at various stages of this research.
\end{acknowledgements}

\bibliographystyle{aa}          
\bibliography{ref}

\begin{appendix} 
\section{The centre of the SNR G42.1-0.2}
\label{apdx:snrcentre}

\begin{figure}[t]
    \centering
    \includegraphics[width=0.48\textwidth]{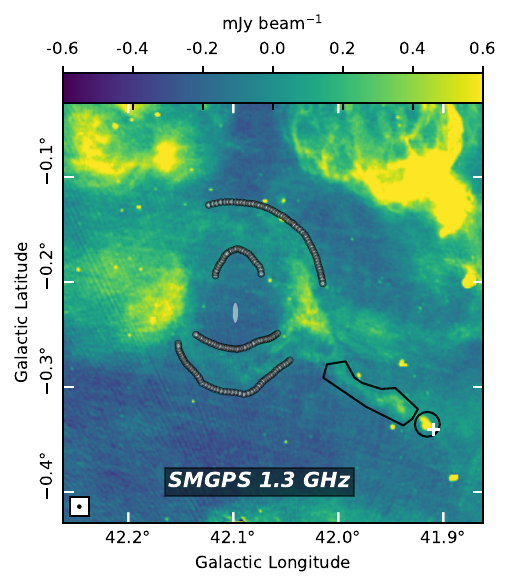}
    \caption{SMGPS image of the SNR G42.1-0.2 and the points used to fit ellipses and estimate the centre of the SNR (marked with gray circles outlined by dark edges).  The white ellipse represents the position of the centre of the SNR estimated using bootstrapping, with its width and height representing the $1\sigma$-uncertainties.  The pulsar PSR\,J1909+0749 is marked with a white cross.  The areas used to calculate the spectral indices of the possible PWN and the cometary tail are marked by a black circle and a black polygon, respectively.  }
    \label{fig:centrepoints}
\end{figure}

In order to quantify the method of determining the centre coordinates of the SNR G42.1-0.2 and the uncertainties, we used a bootstrapping approach.  First, we picked points from the inner and outer edges of the diffuse emission that seemed to clearly mark the extent of the SNR.  This resulted in four arcs representing northern-inner, northern-outer, southern-inner, and southern-outer edges (see Fig.\,\ref{fig:centrepoints}).  Then, on each of the four arcs, 50\% of the points were randomly chosen to fit an ellipse using the \texttt{measure.EllipseModel} class of \texttt{scikit-image} package \citep{2014PeerJ...2..453V}, and the centres of the four ellipses were noted.  This process was repeated for a total of 1000 runs, resulting in 4000 values of centre coordinates.  The mean and the standard deviation of these 4000 values of Galactic coordinates are taken as the centre of the SNR and its uncertainty: $l=42.097\degree\pm 0.005\degree$, $b=-0.229\degree\pm 0.019\degree$.  They are represented as a white ellipse in Fig.\,\ref{fig:centrepoints}.

\section{Surface magnetic field}
\label{apdx:bsurf}

\begin{figure}[t]
    \centering
    \includegraphics[width=0.45\textwidth]{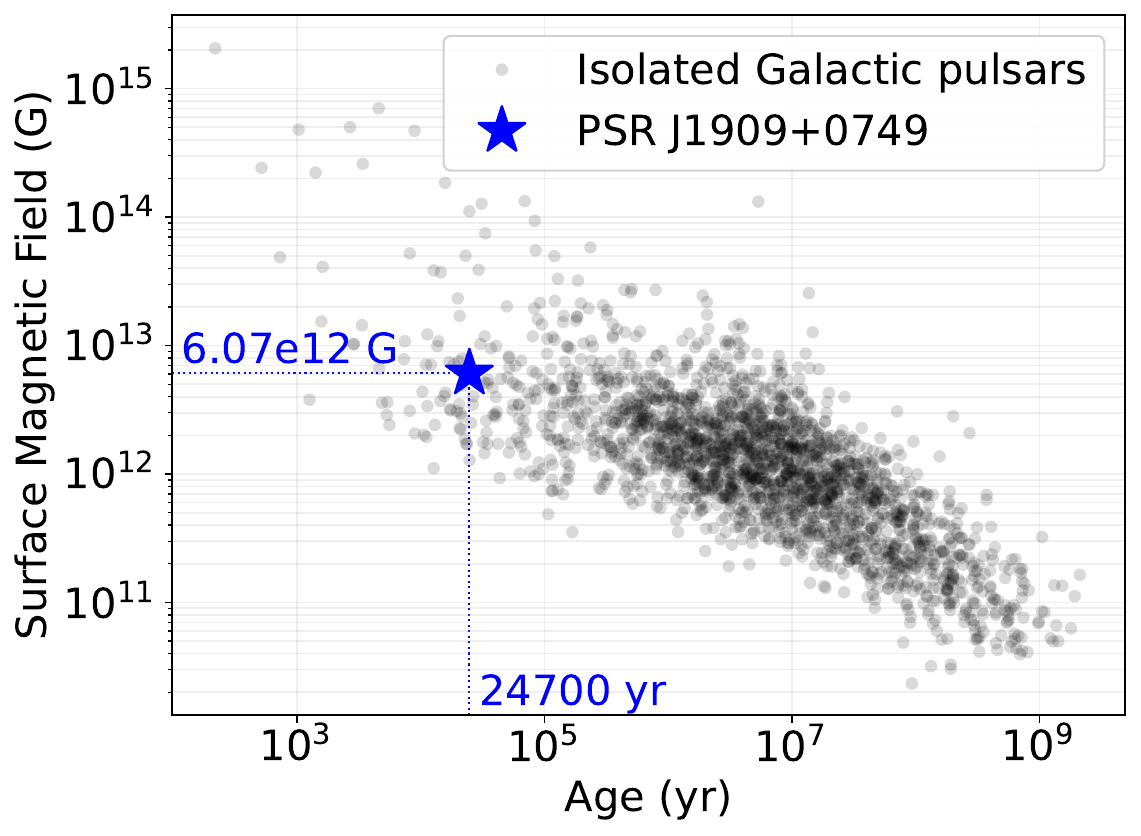}
    \caption{Distribution of the derived surface magnetic field of pulsars as a function of their characteristic age, based on the values from the ATNF pulsar catalogue \citep{2005AJ....129.1993M}.}
    \label{fig:psr_age}
\end{figure}

Figure~\ref{fig:psr_age} shows the surface magnetic field as a function of the age of the pulsar, with PSR\,J1909+0749 marked by a star.  We note that both the age, $\tau_c$, and the magnetic field, $B$, are related quantities derived from the time period, $P$, and its derivative, $\dot{P}$: $\tau_c=P/(2\dot{P})$ and $B=3.2\times 10^{19} (P\dot{P})^{1/2}$~G.  Only isolated Milky Way pulsars with DM-based distances smaller than 25 kpc are shown in this figure.

\section{Pulsar Transverse Velocities}
\label{apdx:ps_tvels}

\begin{figure}[t]
    \centering
    \includegraphics[width=0.45\textwidth]{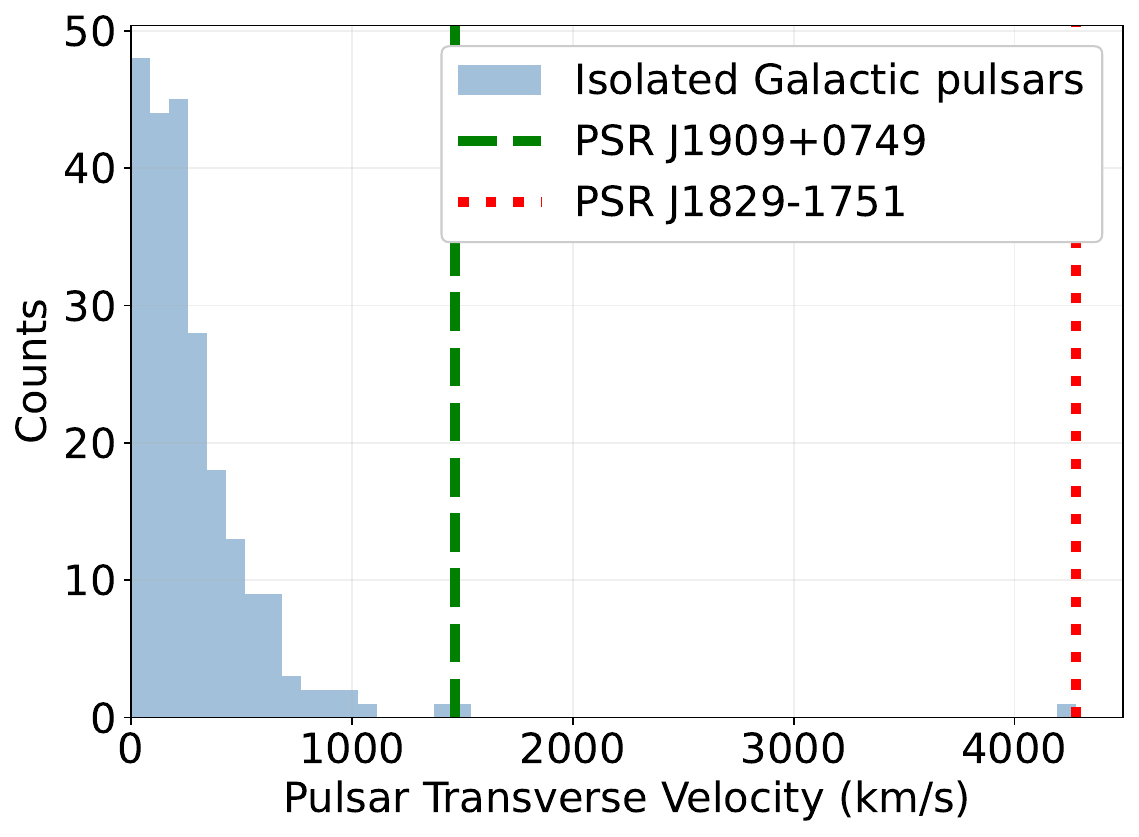}
    \caption{Histogram of the transverse velocities of pulsars derived from the distance and the proper motion values in the ATNF pulsar catalogue \citep{2005AJ....129.1993M}.   PSR\,J1909+0749 has an inferred velocity of about 1460~km\,s$^{-1}$. }
    \label{fig:psr_tvel}
\end{figure}

In Figure~\ref{fig:psr_tvel}, the distribution of the transverse velocities of isolated Milky Way pulsars is shown.  The velocities are calculated using the DM-based distance and the proper motion measurements noted in the ATNF pulsar catalogue \citep{2005AJ....129.1993M}.  Only isolated Milky Way pulsars with DM-based distances smaller than 25 kpc are considered in this plot.  There are two pulsars with absolute values of transverse velocities larger than that of PSR\,J1909+0749.  They are PSR\,J1901+0331 ($1530\pm 170$~km~s$^{-1}$) and PSR\,J1829-1751 ($4300\pm 3700$~km~s$^{-1}$).

\end{appendix} 

\end{document}